\documentstyle[12pt,epsf]{article}
\begin{document}
\title{
{\bf Fast methods for computing the Neuberger Operator}
}

\author{Artan Bori\c{c}i \\
        {\normalsize\it Paul Scherrer Institute}\\
        {\normalsize\it CH-5232 Villigen PSI}\\
        {\normalsize\it Artan.Borici@psi.ch}
        \thanks{Talk given at the ``Interdisciplinary Workshop on
        Numerical Challenges to Lattice QCD'', Wuppertal, 22-24 August 1999}
}

\date{}
\maketitle

\begin{abstract}
I describe a Lanczos method to compute the Neuberger Operator and
a multigrid algorithm for its inversion.
\end{abstract}

\section{Introduction}

Quantum Chromodynamics (QCD) is a theory of strong interactions,
where the chiral symmetry plays a mayor role.
There are different starting points to formulate
a lattice theory with exact chiral symmetry, but
all of them must obey the Ginsparg-Wilson condition
\cite{Ginsparg_Wilson}:
\begin{equation}
\gamma_5 D^{-1} + D^{-1} \gamma_5 = a \gamma_5 {\alpha}^{-1},
\end{equation}
where $a$ is the lattice spacing,
$D$ is the lattice Dirac operator and ${\alpha}^{-1}$ is a local
operator and trivial in the Dirac space.

A candidate is the overlap operator of Neuberger \cite{Neuberger1}:
\begin{equation}
D = 1 - A (A^{\dag}A)^{-1/2}, ~~~~A = M - aD_{W}
\end{equation}
where $M$ is a shift parameter in the range $(0,2)$,
which I have fixed at one and $D_{W}$ is the Wilson-Dirac operator,
\begin{equation}
D_{W} = \frac{1}{2} \sum_{\mu}
[\gamma_{\mu} (\partial_{\mu}^{*} + \partial_{\mu})
            - a\partial_{\mu}^{*}\partial_{\mu}]
\end{equation}
and $\partial_{\mu}$ and $\partial_{\mu}^{*}$ are the nearest-neighbor
forward and backward difference operators, which are covariant, i.e.
the shift operators pick up a unitary 3 by 3 matrix with determinant one.
These small matrices are associated with the links of the lattice and
are oriented positively. A set of such matrices forms a "configuration".
$\gamma_{\mu}, \mu = 1, \ldots, 5$
are 4 by 4 matrices related to the spin of the particle.
Therefore, if there are $N$ lattice points, the matrix is of order
$12N$. A restive symmetry of the matrix $A$ that comes from the
continuum is the so called $\gamma_5-symmetry$ which is the Hermiticity
of the $\gamma_5 A$ operator.

\bigskip
The computation of the inverse square root of a matrix is reviewed
in \cite{Higham}. In the context of lattice QCD there are several
sparse matrix methods, which are developed recently
\cite{Borici,Neuberger2,Edwards_Heller_Narayanan,Bunk,Hern_Jans_Lello}.
I will focus here
on a Lanczos method similar to \cite{Borici}. For a more general case
of functions of matrices I refer to the talk of H. van der Vorst, and
for a Chebyshev method I refer to the talk of K. Jansen, both included in
these proceedings. 

\section{The Lanczos Algorithm}

The Lanczos iteration is known to approximate the spectrum of
the underlying matrix in an optimal way and, in particular,
it can be used to solve linear systems \cite{Golub_VanLoan}.

Let $Q_n = [q_1,\ldots,q_n]$ be the set of orthonormal vectors,
such that
\begin{equation}\label{HQ_QT}
A^{\dag}A Q_n = Q_n T_n + \beta_n q_{n+1} (e_n^{(n)})^T,
~~~~q_1 = \rho_1 b, ~~~~\rho_1 = 1/||b||_2
\end{equation}
where $T_n$ is a tridiagonal and symmetric matrix,
$b$ is an arbitrary vector, and $\beta_n$ a real and positive constant.
$e_m^{(n)}$ denotes the unit vector with $n$ elements
in the direction $m$.

By writing down the above decomposition in terms of the vectors
$q_i, i=1,\ldots,n$ and the matrix elements of $T_n$, I arrive at
a three term recurrence that allows to compute these vectors
in increasing order, starting from the vector $q_1$. This is
the $Lanczos Algorithm$:
\begin{equation}
\begin{array}{l}
\beta_0 = 0, ~\rho_1 = 1 / ||b||_2, ~q_0 = o, ~q_1 = \rho_1 b \\
for ~i = 1, \ldots \\
~~~~v = A^{\dag}A q_i \\
~~~~\alpha_i = q_i^{\dag} v \\
~~~~v := v - q_i \alpha_i - q_{i-1} \beta_{i-1} \\
~~~~\beta_i = ||v||_2 \\
~~~~if \beta_i < tol, ~n = i, ~end ~for \\
~~~~q_{i+1} = v / \beta_i \\
\end{array}
\end{equation}
where $tol$ is a tolerance which serves as a stopping condition.

The Lanczos Algorithm constructs a basis for the Krylov subspace
\cite{Golub_VanLoan}:
\begin{equation}
\mbox{span}\{b,A^{\dag}Ab,\ldots,(A^{\dag}A)^{n-1}b\}
\end{equation}
If the Algorithm stops after $n$ steps, one says that the associated
Krylov subspace  is invariant.

In the floating point arithmetic, there is a danger that
once the Lanczos Algorithm (polynomial) has approximated well some part
of the spectrum,
the iteration reproduces vectors which are rich
in that direction \cite{Golub_VanLoan}.
As a consequence, the orthogonality of the Lanczos vectors is spoiled
with an immediate impact on the history of the iteration: if the
algorithm would stop after $n$ steps in exact arithmetic,
in the presence of round off errors the loss
of orthogonality would keep the algorithm going on.

\section{The Lanczos Algorithm for solving $A^{\dag}A x = b$}

Here I will use this algorithm to solve linear systems, where the loss
of orthogonality will not play a role in the sense that I will use
a different stopping condition.

I ask the solution in the form
\begin{equation}
x = Q_n y_n
\end{equation}
By projecting the original system on to the Krylov subspace I get:
\begin{equation}
Q_n^{\dag} A^{\dag}A x = Q_n^{\dag} b
\end{equation}
By construction, I have
\begin{equation}
b = Q_n e_1^{(n)} / \rho_1,
\end{equation}
Substituting $x = Q_n y_n$ and using (\ref{HQ_QT}), my task is
now to solve the system
\begin{equation}
T_n y_n = e_1^{(n)} / \rho_1
\end{equation}
Therefore the solution is given by
\begin{equation}
x = Q_n T_n^{-1} e_1^{(n)} / \rho_1
\end{equation}

This way using the Lanczos iteration one reduces the size of the
matrix to be inverted. Moreover, since $T_n$ is tridiagonal, one
can compute $y_n$ by short recurences.

If I define:
\begin{equation}
r_i = b - A^{\dag}A x_i, ~~q_i = \rho_i r_i, ~~y_i = \rho_i x_i
\end{equation}
where $i = 1, \ldots$, it is easy to show that
\begin{equation}
\begin{array}{l}
\rho_{i+1} \beta_i + \rho_i \alpha_i + \rho_{i-1} \beta_{i-1} = 0 \\
q_i + y_{i+1} \beta_i + y_i \alpha_i + y_{i-1} \beta_{i-1}  = 0
\end{array}
\end{equation}

Therefore the solution can be updated
recursively and I have the following
{\em Algorithm1 for solving the system $A^{\dag}A x = b$:}
\begin{equation}
\begin{array}{l}
\beta_0 = 0, ~\rho_1 = 1 / ||b||_2, ~q_0 = o, ~q_1 = \rho_1 b \\
for ~i = 1, \ldots \\
~~~~v = A^{\dag}A q_i \\
~~~~\alpha_i = q_i^{\dag} v \\
~~~~v := v - q_i \alpha_i - q_{i-1} \beta_{i-1} \\
~~~~\beta_i = ||v||_2 \\
~~~~q_{i+1} = v / \beta_i \\
~~~~y_{i+1} = - \frac{q_i + y_i \alpha_i + y_{i-1} \beta_{i-1}}{\beta_i} \\
~~~~\rho_{i+1} = - \frac{\rho_i \alpha_i + \rho_{i-1} \beta_{i-1}}{\beta_i} \\
~~~~r_{i+1} := q_{i+1} / \rho_{i+1} \\
~~~~x_{i+1} := y_{i+1} / \rho_{i+1} \\
~~~~if \frac{1}{|\rho_{i+1}|} < tol, ~n = i, ~end ~for \\
\end{array}
\end{equation}

\section{The Lanczos Algorithm for solving $(A^{\dag}A)^{1/2} x = b$}

Now I would like to compute $x = (A^{\dag}A)^{-1/2} b$ and still
use the Lanczos Algorithm. In order to do so I make the following
observations:

Let $(A^{\dag}A)^{-1/2}$ be expressed by a matrix-valued function,
for example the integral formula \cite{Higham}:
\begin{equation}
(A^{\dag}A)^{-1/2} = \frac{2}{\pi} \int_0^{\infty} dt (t^2 + A^{\dag}A)^{-1}
\end{equation}

From the previous section, I use the Lanczos Algorithm to compute
\begin{equation}
(A^{\dag}A)^{-1} b = Q_n T_n^{-1} e_1^{(n)} / \rho_1
\end{equation}

It is easy to show that the Lanczos Algorithm is shift-invariant.
i.e. if the matrix $A^{\dag}A$ is shifted by a constant say, $t^2$,
the Lanczos vectors remain invariant. Moreover, the corresponding
Lanczos matrix is shifted by the same amount.

This property allows one to solve the system
$(t^2 + A^{\dag}A) x = b$ by using the same Lanczos iteration
as before. Since the matrix $(t^2 + A^{\dag}A)$ is better conditioned
than $A^{\dag}A$, it can be concluded that once the
original system is solved, the shifted one is solved too.
Therefore I have:
\begin{equation}
(t^2 + A^{\dag}A)^{-1} b = Q_n (t^2 + T_n)^{-1} e_1^{(n)} / \rho_1
\end{equation}

Using the above integral formula and puting everything together,
I get:
\begin{equation}\label{result}
x = (A^{\dag}A)^{-1/2} b = Q_n T_n^{-1/2} e_1^{(n)} / \rho_1
\end{equation}

There are some remarks to be made here:

a) As before, 
by applying the Lanczos iteration on $A^{\dag}A$, the problem
of computing $(A^{\dag}A)^{-1/2} b$ reduces to the problem of computing
$y_n = T_n^{-1/2} e_1^{(n)} / \rho_1$ which is typically
a much smaller problem than the original one. But since $T_n^{1/2}$
is full, $y_n$ cannot be computed by short recurences.
It can be computed for example by using the full
decomposition of $T_n$ in its eigenvalues and eigenvectors; in fact
this is the method I have employed too, for its compactness and
the small overhead for moderate $n$.

b) The method is not optimal, as it would have been, if one would have
applied it directly for the matrix $(A^{\dag}A)^{1/2}$.
By using $A^{\dag}A$ the condition is
squared, and one looses a factor of two compared to the theoretical case!

c) From the derivation above, it can be concluded
that the system $(A^{\dag}A)^{1/2} x = b$
is solved at the same time as the system $A^{\dag}A x = b$.

d) To implement the result (\ref{result}),
I first construct the Lanczos matrix
and then compute
\begin{equation}
y_n = T_n^{-1/2} e_1^{(n)} / \rho_1
\end{equation}
To compute $x = Q_n y_n$, I repeat the Lanczos iteration.
I save the scalar products, though it is not necessary.

Therefore I have the following
{\em Algorithm2 for solving the system $(A^{\dag}A)^{1/2} x = b$:}
\begin{equation}
\begin{array}{l}
\beta_0 = 0, ~\rho_1 = 1 / ||b||_2, ~q_0 = o, ~q_1 = \rho_1 b \\
for ~i = 1, \ldots \\
~~~~v = A^{\dag}A q_i \\
~~~~\alpha_i = q_i^{\dag} v \\
~~~~v := v - q_i \alpha_i - q_{i-1} \beta_{i-1} \\
~~~~\beta_i = ||v||_2 \\
~~~~q_{i+1} = v / \beta_i \\
~~~~\rho_{i+1} = - \frac{\rho_i \alpha_i + \rho_{i-1} \beta_{i-1}}{\beta_i} \\
~~~~if \frac{1}{|\rho_{i+1}|} < tol, ~n = i, ~end ~for \\
\\
Set ~(T_n)_{i,i} = \alpha_i, ~(T_n)_{i+1,i} = (T_n)_{i,i+1} = \beta_i,
otherwise ~(T_n)_{i,j} = 0 \\
y_n = T_n^{-1/2} e_1^{(n)} / \rho_1
= U_n \Lambda_n^{-1/2} U_n^T e_1^{(n)} / \rho_1 \\
\\
q_0 = o, ~q_1 = \rho_1 b, ~x_0 = o \\
for ~i = 1, \ldots, n \\
~~~~x_i = x_{i-1} + q_i y_n^{(i)} \\
~~~~v = A^{\dag}A q_i \\
~~~~v := v - q_i \alpha_i - q_{i-1} \beta_{i-1} \\
~~~~q_{i+1} = v / \beta_i \\
\end{array}
\end{equation}
where by $o$ I denote a vector with zero entries
and $U_n, \Lambda_n$ the matrices of the egienvectors and eigenvalues
of $T_n$. Note that
there are only four large vectors necessary to store: $q_{i-1},q_i,v,x_i$.

\section{Testing the method}

I propose a simple test: I solve the system $A^{\dag}A x = b$ by
applying twice the $Algorithm2$, i.e. I solve the linear systems
\begin{equation}
(A^{\dag}A)^{1/2} z = b, ~~(A^{\dag}A)^{1/2} x = z
\end{equation}
in the above order. For each approximation $x_i$, I compute the
residual vector
\begin{equation}
r_i = b - A^{\dag}A x_i
\end{equation}

The method is tested for a SU(3) configuration at $\beta = 6.0$
on a $8^316$ lattice, corresponding to an order $98304$ complex
matrix $A$.

In Fig.1 I show the norm of the residual vector decreasing monotonically.
The stagnation of $||r_i||_2$
for small values of $tol$ may come from the accumulation of round off
error in the $64$-bit precision arithmetic used here.

This example shows that the tolerance line is above the residual
norm line, which confirms the expectation that $tol$ is a good
stopping condition of the $Algorithm2$.

\section{Inversion}

Having computed the operator, one can invert it by applying iterative
methods based on the the Lanczos algorithm. Since the operator
$D$ is normal, it turns out that the Conjugate Residual (CR) algorithm is the optimal
one \cite{Borici_thesis}.

In Fig. 2 I show the converegence history of CR on $30$ small $4^4$
lattices at $\beta = 6$. The large number of multiplications with $D_W$
suggests that the inversion of the Neuberger operator is a difficult task
and may bring the complexity of quenched simulations in lattice QCD to
the same order of magnitude to dynamical simulations with Wilson fermions.

Therefore, other ideas are needed.

The essential point is the large number of small eigenvalues of $A$ that make
the computation of $D$ time consuming. Therefore, one may try to project out
these modes and invert them directly \cite{Heller_dubna}.

Also, one may try $5-$dimensional implementations of the Neuberger
operator, such that its condition improves \cite{Neuberger_Pisa}.

I have tried also to reformulate the theory in 5 dimensions by using the
corresponding approximate inversion as a coarse grid solution in a multigrid
scheme \cite{Borici_MG}.

The scheme is tested and the results are shown
in Fig. 2, where the multigrid pattern of the residual norm is clear.
The gain with respect to CR is about a factor $10$.

Note that to invert the ``big'' matrix
I have used the BiCGstab2 algorithm \cite{MGutknecht} which is almost
optimal in most of the cases for the non-normal matrices as its is the matrix
$\cal M$  \cite{Borici_thesis}.

\section{Acknowledgement}

The author would like to thank the organizers of this Workshop for the
kind hospitality at Wuppertal.

\pagebreak

\begin{figure}
\epsfxsize=12cm
\epsfxsize=10cm
\vspace{3cm}
\centerline{\epsffile[100 200 500 450]{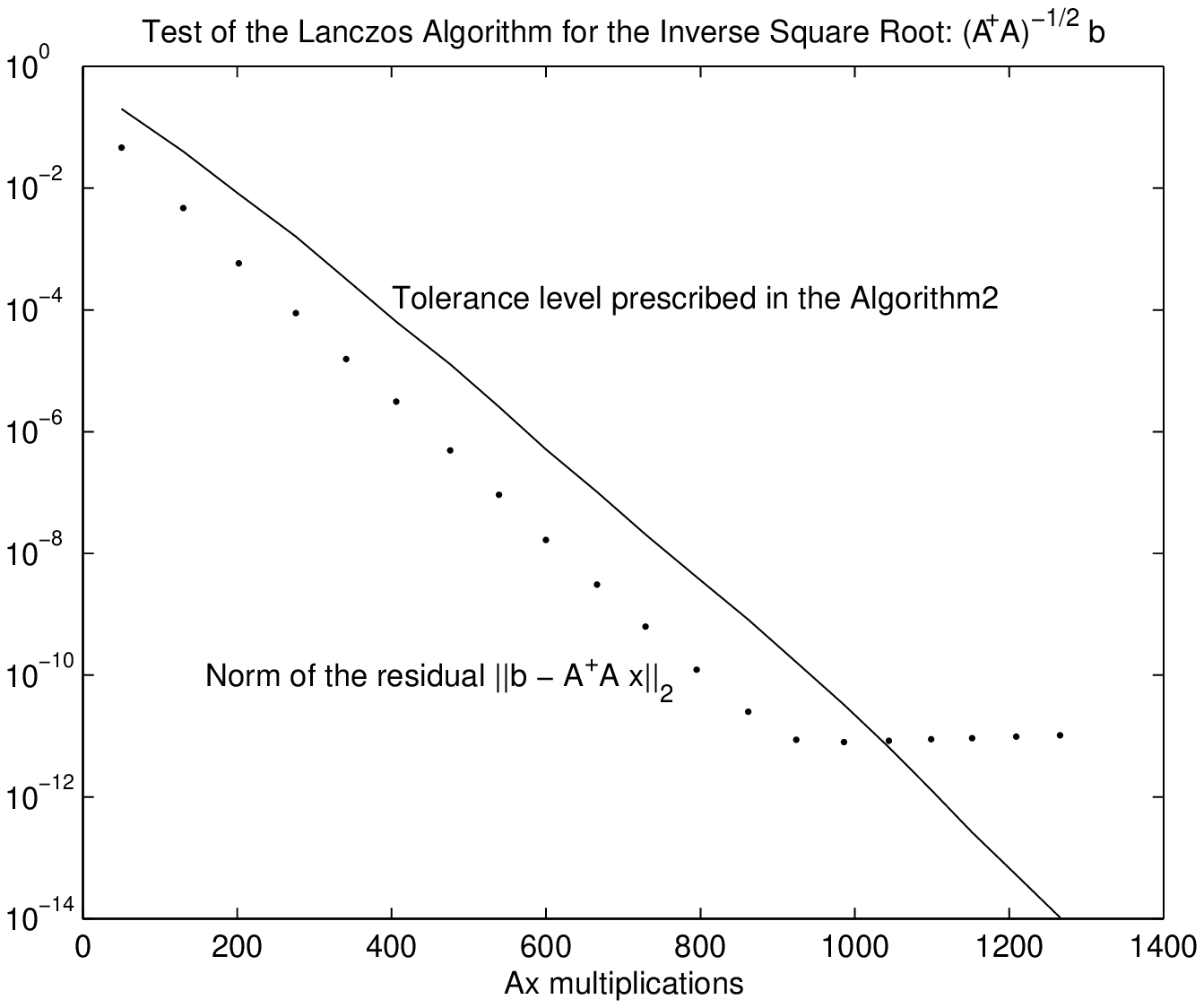}}
\caption{The dots show the norm of the residual vector, whereas
the line shows the tolerance level set by $tol$ in the $Algorithm2$.}
\end{figure}
\begin{figure}
\epsfxsize=12cm
\vspace{3cm}
\centerline{\epsffile[100 200 500 450]{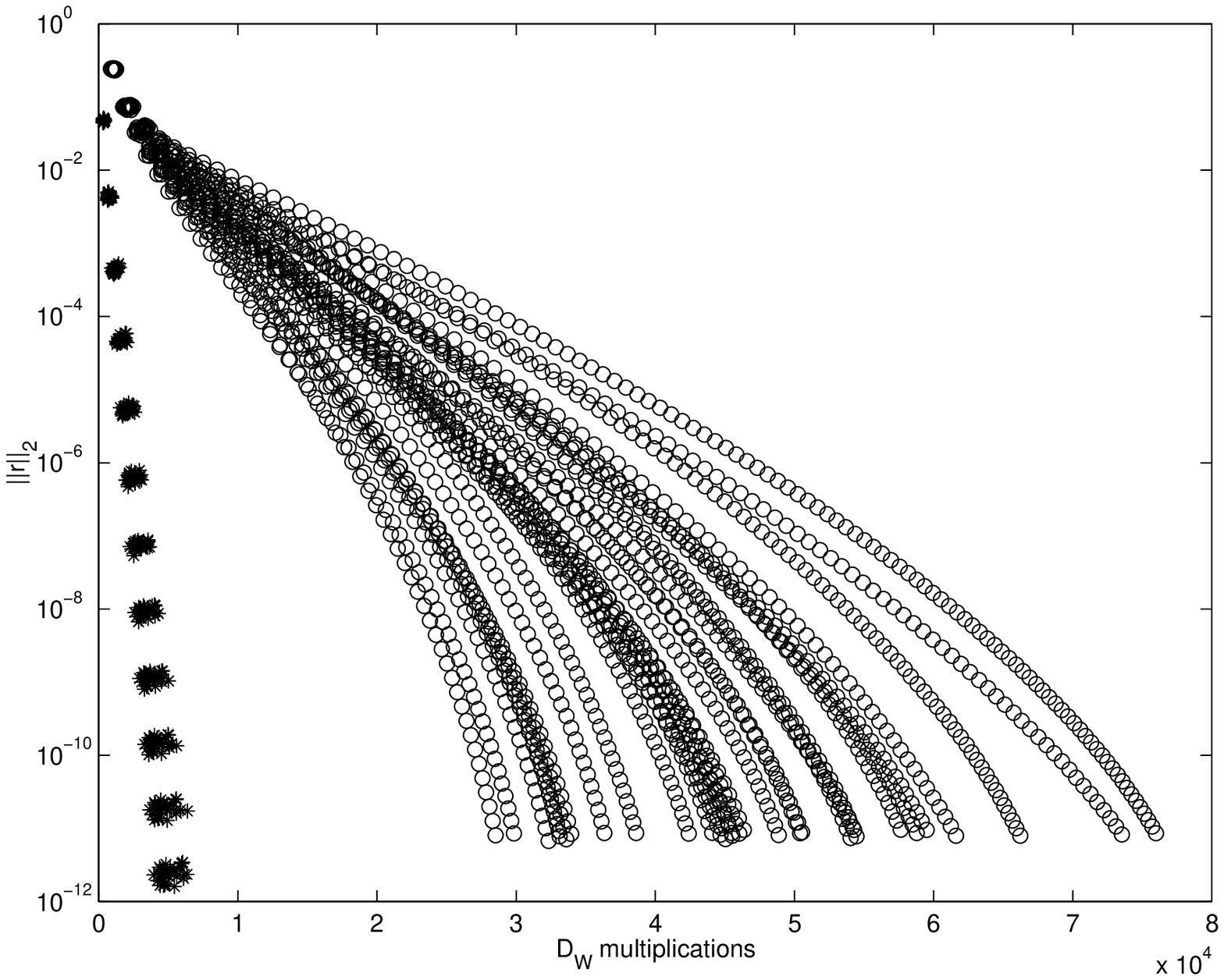}}
\caption{Norm of the residual error vs. the number of $D_W$
multiplications on 30 configurations. Circles stand
for the straightforward inversion with CR and stars for the
multigrid algorithm.}
\end{figure}


\begin{thebibliography}{9}

\bibitem{Ginsparg_Wilson}
         P. H. Ginsparg and K. G. Wilson,
         Phys. Rev. D 25 (1982) 2649.

\bibitem{Neuberger1}
         H. Neuberger,
         Phys. Lett. B 417 (1998) 141,
         Phys. Rev. D 57 (1998) 5417.

\bibitem{Higham}
         N. J. Higham,
         Proceedings of
         "Pure and Applied Linear Algebra: The New Generation",
         Pensacola, March 1993.

\bibitem{Borici}
         A. Bori\c{c}i, Phys.Lett. B453 (1999) 46-53,
         hep-lat/9910045

\bibitem{Neuberger2}
         H. Neuberger,
         Phys. Rev. Lett. 81 (1998) 4060.

\bibitem{Edwards_Heller_Narayanan}
         R. G. Edwards, U. M. Heller and R. Narayanan,
         Nucl.Phys. B540 (1999) 457-471.

\bibitem{Bunk}
         B. Bunk,
         Nucl.Phys.Proc.Suppl. B63 (1998) 952.

\bibitem{Hern_Jans_Lello}
         P. Hernandes, K. Jansen, L. Lellouch,
         these proceedings and hep-lat/0001008.

\bibitem{Golub_VanLoan}
         G. H. Golub and C. F. Van Loan,
         {\it Matrix Computations}, The Johns Hopkins University
         Press, Baltimore, 1989.
         This is meant as a general reference with original
         references included therein.

\bibitem{Borici_thesis}
         A. Bori\c{c}i,
         {\em Krylov Subspace Methods in Lattice QCD},
         PhD Thesis, CSCS TR-96-27, ETH Z\"urich 1996.

\bibitem{Heller_dubna}
         R. G. Edwards, U. M. Heller, J. Kiskis, R. Narayanan,
         hep-lat/9912042.

\bibitem{Neuberger_Pisa}
         H. Neuberger,
         hep-lat/9909042

\bibitem{Borici_MG}
         A. Bori\c{c}i,
         hep-lat/9907003, hep-lat/9909057

\bibitem{MGutknecht}
         M. H. Gutknecht,
         SIAM J. Sci. Comput., 14 (1993) 1020.

\end{thebibliography}
\end{document}